\documentclass[twocolumn,nofootinbib,showpacs,amsmath,amssymb,aps,prd,superscriptaddress]{revtex4}
\usepackage{graphicx}
\usepackage{txfonts}

\usepackage{amsmath}
\usepackage{amssymb}
\usepackage{graphicx}

\usepackage{color}

\usepackage[utf8]{inputenc}
\usepackage{graphicx}
\usepackage{dcolumn}
\usepackage{bm}
\usepackage{txfonts}
\usepackage{color}
\usepackage{comment}
\usepackage{wasysym}

\begin{document}

\newcommand{\lp}{\left(}
\newcommand{\rp}{\right)}
\newcommand{\lb}{\left[}
\newcommand{\rb}{\right]}

\newcommand{\pa}{\parallel}
\newcommand{\pe}{\perp}

\def\bea{\begin{eqnarray}}
\def\eea{\end{eqnarray}}

\newcommand{\Tr}{\mbox{Tr}}
\newcommand{\tr}{\mbox{tr}}
\newcommand{\Det}{\mbox{det}}
\newcommand{\mpl}{M_\mathrm{pl}}
\newcommand{\Dim}[1]{\mbox{dim[$#1$]}}
\newcommand{\del}{\partial}
\newcommand{\ba}{\begin{eqnarray}}
\newcommand{\ea}{\end{eqnarray}}
\newcommand{\be}{\begin{equation}}
\newcommand{\ee}{\end{equation}}
\newcommand{\bk}{{\bf k}}
\newcommand{\bp}{{\bf p}}
\newcommand{\bq}{{\bf q}}
\newcommand{\br}{{\bf r}}
\newcommand{\bx}{{\bf x}}
\newcommand{\om}{\omega}
\newcommand{\Om}{\Omega}
\newcommand{\uA}{\uparrow}
\newcommand{\dA}{\downarrow}
\newcommand{\lA}{\leftarrow}
\newcommand{\rA}{\rightarrow}
\newcommand{\ven}{\varepsilon}
\newcommand{\nn}{\nonumber}
\newcommand{\ket}[1]{\mbox{$\mid\!#1\rangle$}}
\newcommand{\bra}[1]{\mbox{$\langle#1\!\mid$}}
\newcommand{\wh}[1]{\widehat{#1}}
\newcommand{\ex}[1]{\langle #1 \rangle}
\newcommand{\D}{{\cal D}}
\newcommand{\lam}{\bar{\lambda}}
\newcommand{\half}{{1\over 2}}
 
\newcommand{\oo} {\omega}
\newcommand{\al}{\alpha}
\newcommand{\bt}{\beta}
\newcommand{\ga}{\gamma}
\newcommand{\ka}{\kappa}
\newcommand{\ta}{\theta}
\newcommand{\va}{\vartheta}
\newcommand{\vi}{\varphi}
\newcommand{\da}{\delta}
\newcommand{\la}{\lambda}
\newcommand{\za}{\Psi}
\newcommand{\sa}{\sigma}
\newcommand{\en}{\epsilon}
\newcommand{\oa}{\omega}
\newcommand{\Ga}{\Gamma}
\newcommand{\Sa}{\Sigma}
\newcommand{\Ta}{\Theta}
\newcommand{\Da}{\Delta}
\newcommand{\Oa}{\Omega}
\newcommand{\La}{\Lambda}

\newcommand{\R}{\hat{R}}
\newcommand{\h}{\mathcal{H}}
\newcommand{\OA}{\Omega_A}
\newcommand{\e}{\rm e}
\newcommand{\g}{\tilde{g}}
\newcommand{\RR}{\tilde{R}}
\newcommand{\cR}{{\cal R}}

\newcommand{\F}{\OA+\phi}
\newcommand{\Fp}{\lp\OA+\phi\rp}
\newcommand{\f}{f^{(0)}_{,R\R}}
\newcommand{\Q}{f^{(0)}_{,\hat Q}}
\newcommand{\Frr}{f^{(0)}_{,RR}}

\newcommand{\question}{{\color{red}(?)}}
\newcommand{\Mplsq}{M_\mathrm{pl}^2}
\newcommand{\Mplsqh}{\frac{M_\mathrm{pl}^2}{2}}
\newcommand{\dd}{\mathrm{d}}

\title{Screening mechanisms in hybrid metric-Palatini  gravity}

\author{Marcelo Vargas dos Santos}
\email{mvsantos@protonmail.com}
\affiliation{Instituto de F\'isica - Universidade Federal do Rio de Janeiro. Av. Athos da Silveira Ramos, 149. Rio de Janeiro - RJ Brazil}
\affiliation{Departamento de Astronomia, Observat\'orio Nacional, 20921-400, Rio de Janeiro, RJ, Brasil}

\author{Jailson S. Alcaniz}
\email{alcaniz@on.br}
\affiliation{Departamento de Astronomia, Observat\'orio Nacional, 20921-400, Rio de Janeiro,  RJ, Brasil}
\affiliation{Physics Department, McGill University, Montreal QC, H3A 2T8, Canada}
 
\author{David F. Mota}
\email{d.f.mota@astro.uio.no}
\affiliation{Institute of Theoretical Astrophysics, University of Oslo, Postbox 1029, 0315 Oslo, Norway}

\author{Salvatore Capozziello}
\email{capozziello@na.infn.it}
\affiliation{Dipartimento di Fisica, Universit\`a di Napoli ''Federico II'', Via Cinthia, I-80126, Napoli, Italy,}
\affiliation{Istituto Nazionale di Fisica Nucleare (INFN), Sez. di Napoli, Via Cinthia, Napoli, Italy,}
\affiliation{Gran Sasso Science Institute, Via F. Crispi 7,  I-67100, L'Aquila, Italy,}
\affiliation{Tomsk State Pedagogical University, ul. Kievskaya, 60, 634061 Tomsk, Russia.}

\begin{abstract}
	We investigate the efficiency of screening mechanisms in the hybrid metric-Palatini  gravity. The value of the field is computed around spherical bodies embedded in a background of constant density. We find  a thin shell condition for the field  depending  on  the background field value. 
In order to quantify how the thin shell effect is relevant,  we analyze how it behaves in the neighborhood of different  astrophysical objects (planets, moons or stars). We find that the condition is very well satisfied except only for some peculiar objects. Furthermore we establish bounds on the model using data from solar system experiments such as the spectral deviation measured by the Cassini mission and the stability of the Earth-Moon system, which gives the best constraint to date on $f(R)$ theories. 
These bounds contribute to fix the range of viable hybrid  gravity models.
\end{abstract}
\date{\today}

\maketitle

\section{Introdution}

The discovery of the accelerated expansion of the Universe \citep{Riess1998AJ....116.1009R,Perlmutter1999ApJ...517..565P} brings to cosmology one of the most remarkable puzzles because standard matter  cannot act as engine for such a phenomenon. The straightforward solution is to search for an exotic fluid, dubbed as { dark energy}, capable of giving rise to observed cosmic  acceleration.  Another  solution is to modify or extend the theory of general relativity (GR) in order to explain geometrically the phenomenon. This has been done in recent years and leads to numerous theories of modified gravity where curvature or torsion invariants, or scalar fields  can be  considered as sources  into the effective energy-momentum tensor in the right-hand side of the field equations (see, e.g.,   \citep{OurRept, Padilla, Odintsov,mog1,mog2,Santos:2007bs,mog3,vasilis,Pires:2010fv, teleparallel}).

The main challenge for  modified gravity theories is the measurements of the gravitational strength on Earth and in the solar system~ \citep{Will2006LRR.....9....3W},  where the predictions of GR have been confirmed with great precision. A viable solution to this issue  is to take advantage  of the  so-called screening mechanism, which restores GR in the solar system. In other words, the effects of any modified gravity have to start to work at larger (infra-red)  scales than those where the weak filed limit of GR works very well.  Screening mechanisms \citep{Khoury2010}
are usually triggered by large local matter density or space-time curvature and lead to a convergence of the gravitational strength to its value predicted by GR at local scales. For scalar-tensor gravity several possible screening mechanisms have been discussed  (see, e.g., \citep{Khoury2010,Koivisto:2012za,Joyce2014}). The philosophy essentially consists in considering scalar-field couplings and self-interaction potentials that regulate the strength of the gravitational interaction according to the scale.  

Among the various possibilities,  $f(R)$ gravity is a viable mechanism to generate the  speeding up expansion for primordial cosmic  inflation \citep{Starobinsky:1979ty} and late-time acceleration \cite{Cap2002}. The approach consists in the straightforward possibility to extend GR by considering generic functions of the Ricci scalar $R$ in  the Einstein-Hilbert Lagrangian instead of only the linear action in $R$. Two different variational approaches are usually applied to this class of extended  theories of gravity, namely, the metric and the Palatini formalisms. In the former case,  the connections are assumed to be the Christoffel symbols and the variation of the action is taken with respect to the metric, whilst in the latter the metric and the affine connections are regarded as independent fields, such that the variation is taken with respect to both. As it is well known, these approaches lead to  different equations of motion, being equivalent only in the case of a linear action (GR). However, some shortcomings come out both in metric and Palatini approaches and none of them  is completely free of problems when addressing the dynamics of the Universe at any extragalactic and cosmological scale \cite{OurRept}.

Recently, a new class of extended theories of gravity, consisting of the superposition of the metric Einstein-Hilbert Lagrangian with an $f(R)$ term constructed \`a la Palatini has been proposed in Ref.~\cite{Harko:2011nh,Capozziello:2012ny}. Using the equivalent scalar-tensor representation, it can be shown that a theory, which can also be formulated in terms of the quantity $X \equiv T + R$, where $T$ and $R$ are the traces of the energy-momentum and Ricci tensors, respectively, is able to modify the cosmological large-scale structure without affecting the Solar System dynamics. Such results have motivated a number of analysis on this class of theories. Cosmological consequences 
of the so-called  {\it hybrid metric-Palatini gravity}, including criteria to obtain cosmic acceleration~\cite{Capozziello:2012ny}, dynamical solutions~\cite{Carloni:2015bua}, the dark matter problem~\cite{Capozziello:2012qt}, among others~\cite{Lima:2014aza,Azizi:2015ina,Santos:2016tds,Fu:2016szo,pala}, have been investigated. However, the main conceptual reason for introducing hybrid gravity is the following. As discussed in detail in \cite{Capozziello:2015lza}, if $f(R)$ gravity is represented  in the scalar-tensor form, i.e. in a Brans-Dicke-like representation, one obtains that the Brans-Dicke parameter is $\omega_{BD}=0$ for the metric approach and $\omega_{BD}=-3/2$ for the Palatini approach (see below). Both of them are incompatible with the Solar system constraints, so it seems that any straightforward extension of GR cannot be compared with celestial dynamics because the original Brans-Dicke theory indicates that $\omega_{BD}\rightarrow \infty$. The shortcoming is overcome assuming that the standard GR part of the action, i.e. $R$, is metric, while the further degrees of freedom of the gravitational field, i.e. $f(R)$,  are Palatini. In this sense, the connections acquire a dynamical role and cure the shortcomings of both metric and Palatini representations. In fact,  the scalar-field representation of hybrid gravity, as we will see, can be easily compared to GR because the scalar field derived from the Palatini part has a clear dynamical role consisting in a kinetic and a potential components. Another important motivation comes from galactic dynamics. It is well-known that  that the weak-field limit of any analytic $f(R)$ model gives rise to Yukawa-like corrections into the Newtonian potential.
This result is useful to reproduce the rotation curve of galaxies and the galactic cluster dynamics without assuming huge amounts of dark matter\cite{vincenzo,Napolitano}. Despite of this good feature,  the correction parameter is fixed by the theory and it is difficult to match the observations in a realistic way. As discussed in \cite{essay}, the weak field limit of hybrid gravity allows to overcome  this shortcoming because the correction parameter is related to the dynamical scalar field and then depends on the boundary conditions of the self-gravitating system that one is considering.
We refer the reader to ~\cite{Capozziello:2015lza} for a  review on the motivations for introducing hybrid gravity. 

In this paper, we  investigate the efficiency of screening mechanisms in the hybrid metric-Palatini $f(X)$ gravity. We compute the value of the field around spherical bodies embedded in a background of constant density and impose bounds on the model using data from solar system experiments, such as the spectral deviation measured by the Cassini mission and the stability of the Earth-Moon system. In Sec. II,  we sketch  the hybrid metric-Palatini formalism. The scalar-tensor representation of this class of extended gravity theories is discussed in Sec. III. The thin shell effect is studied in detail in Sec. IV, where some numerical values are derived for the solar system planets and stars that host exoplanets. The behaviour of $f(X)$ gravity in the neighbours of astrophysical bodies is discussed  in Sec. V. Using data from the Cassini mission and the bound  conditions of  the Earth-Moon system, we derive stringent bounds for these class of models. A summary of the results and a final discussion is reported  in Sec. VI.

\section{The hybrid metric-Palatini $f(X)$ gravity}

The action  of hybrid metric-Palatini gravity  can be  written as \cite{Harko:2011nh,Capozziello:2012ny}
\begin{equation}\label{eq:S_hybrid}
S = \Mplsqh\int \dd^4 x \sqrt{-g} \left[ R + f(\R)\right] + S_m(g_{\mu\nu},\Psi),
\end{equation}
where $S_m(g_{\mu\nu},\Psi)$ is the matter action, $M_\mathrm{pl}$ is the Planck mass, $R$ is
the Ricci scalar (in the metric formalism) and $\R \equiv g^{\mu\nu}\R_{\mu\nu} $ is
the Ricci curvature scalar in the Palatini formalism. Such a  Ricci curvature tensor is defined in terms of an independent connection ($\hat{\Gamma}^\alpha_{\mu\nu}$) as
\begin{equation}
\R_{\mu\nu} \equiv \hat{\Gamma}^\alpha_{\mu\nu\prime\alpha}
       - \hat{\Gamma}^\alpha_{\mu\alpha\prime\nu} +
\hat{\Gamma}^\alpha_{\alpha\lambda}\hat{\Gamma}^\lambda_{\mu\nu} -
\hat{\Gamma}^\alpha_{\mu\lambda}\hat{\Gamma}^\lambda_{\alpha\nu} ,\label{r_def}
\end{equation}
Varying the action \eqref{eq:S_hybrid} with respect to the metric, we obtain the following gravitational field equations  
\begin{equation}
\label{efe} G_{\mu\nu} +
F(\R)\R_{\mu\nu}-\frac{1}{2}f(\R)g_{\mu\nu} = \frac{T_{\mu\nu}}{\Mplsq} ,
\end{equation}
where $F(\R):=df/d\R$ and the matter energy-momentum tensor is defined as
 \begin{equation}\label{memt}
 T_{\mu\nu} \equiv -\frac{2}{\sqrt{-g}} \frac{\partial(\sqrt{-g}\mathcal{L}_m)}{\partial g^{\mu\nu}}.
 \end{equation}
Varying the action with respect to the independent connection, $\hat{\Gamma}^\alpha_{\mu\nu}$, we find that the solution of the equations of motion is such that $\hat{\Gamma}^\alpha_{\mu\nu}$ is compatible with the metric $\hat{g}_{\mu\nu} = F(\R)g_{\mu\nu}$, conformally related to the physical metric by a conformal factor $F(\R) \equiv df(\R)/d\R$ (see \cite{OurRept} for details). This implies that
\be
\label{ricci} \R_{\mu\nu} = R_{\mu\nu} +
\frac{3}{2}\frac{F(\R)_{\prime\mu}F(\R)_{\prime\nu}}{F^2(\R)}
  - \frac{\nabla_\mu F(\R)_{\prime\nu}}{F(\R)} -
\frac{g_{\mu\nu} \nabla^2 F(\R)}{2F(\R)}. \ee
Taking the trace of Eq. \eqref{efe}, we find that the relation between the Ricci curvature scalar $R$, in metric formalism,  and the  curvature $\R$, in the Palatini formalism,  is given by \begin{equation}\label{trace} F(\R)\R -2f(\R) = R + \frac{T}{\Mplsq} \equiv X .
\end{equation}
Therefore if the form of $f(\R)$ allows analytic solutions, then $\R$ can be expressed algebraically in terms of $X$. The variable $X$ quantifies how much the theory deviates from GR, which gives the trace equation $R = -T/\Mplsq$. Indeed, the field Eq. \eqref{efe} can be expressed  in terms of the metric and $X$ as
\begin{eqnarray} \label{efex} 
G_{\mu\nu} = F'(X)\nabla_{\mu}X_{\prime\nu} - F(X)R_{\mu\nu} + \nonumber \\ + \frac{1}{2}\lb f(X) + F'(X) \nabla^2 X + F''(X)\lp \partial X\rp^2 \rb
g_{\mu\nu} + \nonumber \\ + \lb F''(X) - \frac{3}{2}\frac{\lp
F'(X)\rp^2}{F(X)}\rb X_{\prime\mu}X_{\prime\nu} + \frac{T_{\mu\nu}}{\Mplsq}, \end{eqnarray}
whose trace gives
\begin{eqnarray} \label{trace2} F'(X)\nabla^2 X + \lb
F''(X)-\frac{\lp F'(X)\rp^2}{2F(X)}\rb \lp \partial
X\rp^2+ \nonumber \\+
 \frac{X + 2f(X)-F(X)R}{3} = 0  ,
\end{eqnarray} 
and the relation between $R$ and $\R$ finally reduces to    
\begin{equation}\label{ricciscalar} \R(X) =
R+\frac{3}{2}\lb \lp\frac{F'(X)}{F(X)}\rp^2-2\frac{\nabla^2
F(X)}{F(X)}\rb , \end{equation} 
which is obtained by contracting Eq. \eqref{ricci}.

\section{Scalar-tensor representation}

As in the pure metric and Palatini cases \cite{Olmo:2005zr,Olmo:2005hc}, the action \eqref{eq:S_hybrid} can be turned into that of a
scalar-tensor theory by introducing an auxiliary field $\chi$. The new action is given by
\begin{equation}\label{eq:S_scalar0}
S= \Mplsqh\int \dd^4 x \sqrt{-g} [R + f(\chi)+f_\chi(\R-\chi)] + S_m(g_{\mu\nu},\Psi) \ ,
\end{equation}
where the sub-index $\chi$ denotes the derivative with respect to the field $\chi$. Varying it with respect to $\chi$, we find that $f_{\chi\chi}(\R-\chi)=0$, which means that it is equivalent to the action \eqref{eq:S_hybrid} since $\R=\chi$ for $f_{\chi\chi}\neq0$. Defining a field as $\phi\equiv f_\chi$ and its potential as $U(\chi)= \chi f_\chi - f(\chi)$, the action \eqref{eq:S_scalar0} becomes
\begin{equation}\label{eq:S_scalar1}
S = \Mplsqh\int \dd^4 x \sqrt{-g} (R + \phi\R-U(\chi)) + S_m(g_{\mu\nu},\Psi).
\end{equation}
Varying the above expression with respect to the metric, the scalar field and the independent connection leads to the field equations
\begin{subequations}
\begin{equation}
R_{\mu\nu} + \phi \R_{\mu\nu} - \frac{1}{2}\left(R+\phi\R-U(\phi)\right)g_{\mu\nu} = \frac{T_{\mu\nu}}{\Mplsq} ,
\label{eq:var-gab}\\
\end{equation}
\begin{equation}
\R = \frac{dU}{d\phi} \label{eq:var-phi}  , \\
\end{equation}
\begin{equation}
\hat{\nabla}_\alpha\left(\sqrt{-g}\phi g^{\mu\nu}\right)=0  , \label{eq:connection}\
\end{equation}
\end{subequations}
respectively. 

The solution of Eq. \eqref{eq:connection} implies that the independent connection is the Levi-Civita connection of a metric $\hat{g}_{\mu\nu}=\phi g_{\mu\nu}$. Therefore, the relation \eqref{ricci} can now be rewritten as
\begin{equation}\label{eq:conformal_Rmn}
\R_{\mu\nu}=R_{\mu\nu}+\frac{3}{2\phi^2}\partial_\mu \phi \partial_\nu \phi-\frac{1}{\phi}\left(\nabla_\mu
\nabla_\nu \phi+\frac{1}{2}g_{\mu\nu}\nabla^2 \phi\right) \ ,
\end{equation}
which can be used in the action \eqref{eq:S_scalar1} to eliminate the independent connection and  obtain the following scalar-tensor representation (which belongs to the ``Algebraic Family of Scalar-Tensor Theories'') \cite{Koivisto:2009jn}:
\begin{eqnarray}\label{eq:S_scalar2}
S &=& \Mplsq\int \dd^4 x \sqrt{-g} \left[ \frac{1}{2}(1+\phi)R +\frac{3}{2\phi}(\partial \phi)^2
- U(\phi)\right] + \nonumber \\ &+& S_m(g_{\mu\nu},\Psi) \ .
\end{eqnarray}
Using Eqs. \eqref{eq:var-gab}, \eqref{eq:var-phi} and \eqref{eq:conformal_Rmn}, the metric field equations can be
written as
\begin{eqnarray}
(1+\phi) R_{\mu\nu} &=& \frac{1}{\Mplsq}\left(T_{\mu\nu}-\frac{1}{2}g_{\mu\nu} T\right) + \frac{1}
{2}g_{\mu\nu}\left[U(\phi) + \nabla^2 \phi\right] + \nonumber \\ &+& \nabla_\mu\nabla_\nu\phi-\frac{3}{2\phi}\partial_\mu \phi
\partial_\nu \phi \ \label{eq:evol-gab} ,
\end{eqnarray}
or, equivalently, as
\begin{eqnarray}\label{einstein_phi}
(1+\phi)G_{\mu\nu} &=& \frac{T_{\mu\nu}}{\Mplsq} + \nabla_\mu\nabla_\nu\phi - g_{\mu\nu}\nabla^2 \phi - \frac{3}{2\phi}\nabla_\mu\phi \nabla_\nu\phi + \nonumber \\ &+& \frac{3}{4\phi}(\nabla\phi)^2 g_{\mu \nu}-
\frac{1}{2}U(\phi)g_{\mu\nu},
\end{eqnarray}
which clearly show that the spacetime curvature is sourced by both the matter and the scalar field.

As discussed above, the scalar field equation can be manipulated in two different ways that illustrate how the hybrid models combine physical features eliminating the shortcoming of Brans-Dicke theory in both metric and Palatini formalism, being  $\omega_{BD}=0$ for metric approach and $\omega_{BD}=-3/2$ for palatini approach for scalar-tensor models \cite{Capozziello:2015lza}. First, tracing Eq.~(\ref{eq:var-gab}) with $g^{\mu\nu}$,  we find $-R-\phi\R + 2U(\phi) = T/\Mplsq$, and using Eq.~(\ref{eq:var-phi}), it takes the following form:
\begin{equation}\label{eq:phi(X)}
X \equiv R + \frac{T}{\Mplsq} = 2U(\phi)-\phi \frac{dU}{d\phi}.
\end{equation}
Similarly to the Palatini ($\omega_{BD}=-3/2$) case, this equation tells us that the field $\phi$ can be expressed as an algebraic function of the scalar $X$, i.e., $\phi=\phi(X)$. In the pure Palatini case, however, $\phi$ is just a function of $T$. The right-hand side of Eq.~\eqref{eq:evol-gab}, therefore, besides containing new matter terms associated with the trace $T$ and its derivatives, also contains the curvature $R$ and its derivatives. Thus, this theory can be seen as a higher-derivative theory in both  matter and  metric fields. However, such an interpretation can be avoided if $R$ is replaced in Eq. \eqref{eq:phi(X)} by the relation 
\begin{equation}
R = \R + 3\frac{\nabla^2 \phi}{\phi}-\frac{3}{2}\left(\frac{\partial \phi}{\phi}\right)^2
\end{equation}
together with $\R = dU/d\phi$. One then finds that the scalar field is governed by the second-order evolution equation that becomes
\be\label{eq:evol-phi}
-\nabla^2 \phi + \frac{1}{2\phi}(\partial \phi)^2 + \frac{\phi}{3} \lb 2U(\phi)-(1+\phi)\frac{dU}{d\phi} \rb = \frac{\phi}{3\Mplsq}T ,
\end{equation}
which is an effective Klein-Gordon equation. This last expression shows that, unlike the Palatini ($\omega_{BD}=-3/2$) case, the scalar field is dynamical. The theory is therefore not affected by the microscopic instabilities that arise in Palatini
models with infrared corrections \cite{Olmo:2011uz}.

Finally, we can perform a conformal transformation into the Einstein frame. The conformal rescaling we need is given by
\begin{equation}
	g_{\mu\nu} \rightarrow \g_{\mu\nu} = A^2(\phi)g_{\mu\nu} = \frac{g_{\mu\nu}}{1+\phi} , 
\end{equation}
and the Einstein frame action then becomes
\begin{eqnarray}\label{eq:S_scalar3}
S &=& \Mplsq\int \dd^4 x \sqrt{-g} \left[ \frac{1}{2}R + \frac{3}{2\phi}\frac{\g^{\al\bt}\phi_{\prime\al}\phi_{\prime\bt}}{( 1 + \phi )^2} - V(\phi)\right] + \\ &+& S_m\left(A^2(\phi)\g_{\mu\nu},\Psi\right) \ .
\end{eqnarray}
where $V(\phi) = U(\phi)/A^4(\phi)$. This can be further put into its canonical form by introducing the rescaled field $\vi$ as
\begin{equation}
\phi = -\tanh^2 \lp \frac{\vi}{2\sqrt{3}}\rp \simeq - \frac{\vi^2}{12}, 
\label{new_field}
\end{equation}
and the final action becomes
\begin{eqnarray}\label{eq:S_scalar4}
S &=& \Mplsq\int d^4 x \sqrt{-\g} \left[ \frac{1}{2}\RR - \frac{1}{2}(\tilde{\partial}\vi)^2 - V(\vi)\right] + \\ &+& S_m\left(A^2(\psi)\g_{\mu\nu},\Psi\right) \ .
\end{eqnarray}
This is a scalar-tensor theory action with a quadratic conformal factor
\begin{equation}
	A(\vi) = \lb 1 - \tanh^2\lp \frac{\vi}{2\sqrt{3}}\rp\rb^{-2} \simeq 1 + \frac{\vi^2}{6}\;,
\end{equation}
which gives the following dynamical equation for the scalar field
\begin{equation}
	\square^2 \vi = V'_{eff}(\vi)\;,
\label{kgordon}
\end{equation}
where the effective potential is given by
\begin{equation}
	V_{eff}(\vi) = V(\vi) - (A(\vi)-1)\tilde{T}\;.
\end{equation}
For a pressureless matter field it becomes
\begin{equation}
	V_{eff}(\vi) = V(\vi) + (A(\vi)-1)\rho = V(\vi) + \frac{\vi^2}{6}\rho,
\end{equation}
The vacuum theory then becomes a canonical scalar-tensor  theory with a very specific potential (stemming out from the original function $f(\R)$ in the Einstein frame). 

With these considerations in mind, we can deal with the screening mechanism for hybrid gravity under the same standard of scalar-tensor theories.

\section{The screening mechanism}

A reliable screening mechanism is certainly one of the most important features that any modified theory of gravity has to satisfy to be physically consistent. Such mechanism ensures that a given model is in accordance with the local observations, such as the solar system, exoplanetary systems or galaxy bounds \cite{Khoury2010}. It arises from the fact that  non-minimum couplings between  the gravitational scalar field and the matter fields give rise o to  fifth force effects  depending on the environment physical properties. In other words, the screening mechanism is related to the fact that the Mach principle is fully taken into account but GR must be recovered to be in agreement with observations (see  in \cite{Libro} for a discussion).

\begin{table*}[]
\centering
\caption{The thin shell parameters for the Solar System planets.}
\label{planets_thinshell}
\begin{tabular}{|c|c|c|c|c|c|c|c|c|}
\hline Planet & Mercury & Venus & Earth & Mars & Jupiter & Saturn & Uranus & Neptune \\ \hline \hline 
 $|\varepsilon|$ & $1.6\times 10^{-5}$ & $9.9\times 10^{-5}$ & $2.9 \times 10^{-4}$ & $3 \times 10^{-3}$ & $1.9\times 10^{-5}$ & $ 4.6\times 10^{-2}$ & $1.2\times 10^{-2}$ & $8\times 10^{-3}$ \\ \hline 
\end{tabular}
\end{table*}

\begin{figure*}[tbp]
\includegraphics[width=\columnwidth]{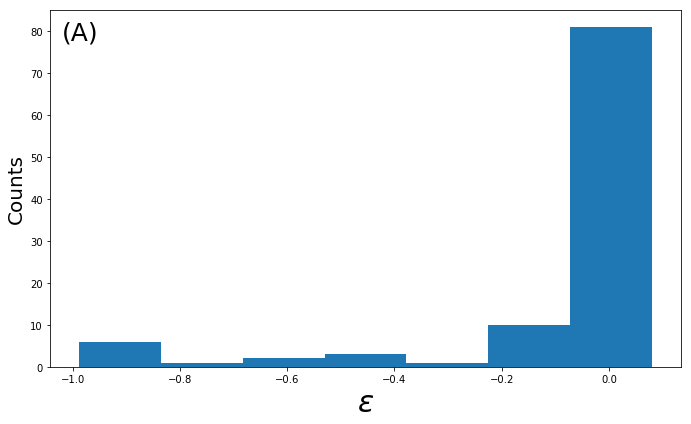}~~~~~~~~~~
\includegraphics[width=\columnwidth]{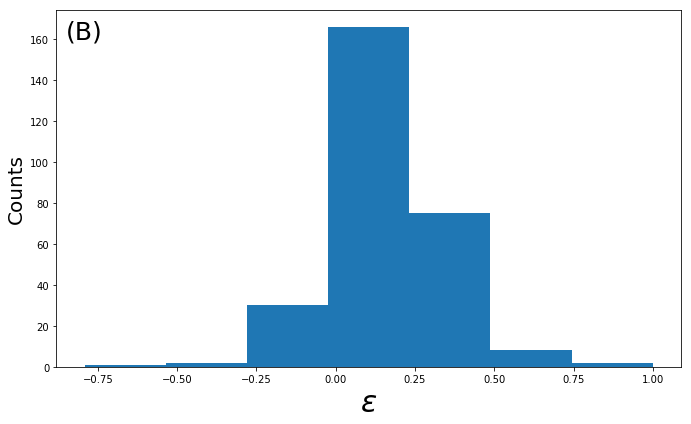}
\caption{Frequency of the thin shell parameter for the solar system moons (A) and stars which hosts planets (B).} \label{hists}
\end{figure*}

\subsection{Spherical solution}

Let us consider a spherically dense body that is embedded in a homogeneous background. Inside this body, the matter density $\rho_c$ is also a constant. i.e., $\rho(r)$ is given by\begin{equation}
	\rho(r) = \left\{ \begin{array}{cc}
          \rho_c, & r < {\cal R} \\
          \rho_b, & r > \cR
          \end{array}, \right.
\end{equation}
where $\cR$ indicates the physical radius of the body. 
In spherical coordinates, the dynamical equation \eqref{kgordon} reduces to
\begin{equation}
	\frac{1}{r^2}\frac{\dd}{\dd r}\left[r^2\frac{\dd\vi}{\dd r}\right] = V'(\vi) + \frac{\rho}{3}\vi.
\end{equation}
In the inner region ($r < \cR$), the density is much higher than the derivative of the potential ($\rho_c \gg V'$) i.e.,
\begin{equation}
	\frac{1}{r^2}\frac{\dd}{\dd r}\left[r^2\frac{\dd\vi}{\dd r}\right] \simeq \frac{\rho_c}{3}\vi,
\end{equation}
whose solution is
\begin{equation}
	\frac{\vi(r<\cR)}{\vi_b} = A \frac{\cR}{r}\sinh m_c r\;,
\end{equation}
where $m_c^2 = \rho_c/3$, $A$ is a constant and $\vi_b$ is the background value of $\vi$. We also have used that $\R \equiv R$ at the minimum of the potential.
In the outer region ($r>\cR$) we expand the potential around the minimum
\begin{equation}
	\frac{1}{r^2}\frac{\dd}{\dd r}\left[r^2\frac{\dd\vi}{\dd r}\right] \simeq m_b^2(\vi-\vi_b),
\end{equation}where $m_b^2 = V''_{eff}(\vi_b)\approx-f_R/3f_{RR}$ is the the background mass. The solution is\begin{equation}
	\frac{\vi(r>\cR)}{\vi_b} = 1 - B\frac{\cR}{r}e^{-m_b (r-\cR)},
\end{equation}
where $B$ is a constant. Imposing that $\vi(r=\cR^-) = \vi(r=\cR^+)$ and $\vi'(r=\cR^-) = \vi'(r=\cR^+)$ as boundary conditions we find the values of $A$ and $B$
\begin{subequations}
\begin{equation}
	A = \frac{1+m_b {\cR}}{(m_c {\cR}) \cosh (m_c {\cR}) + (m_b {\cR}) \sinh (m_c {\cR})}\;,
\end{equation}
\begin{equation}
    B = \frac{(m_c {\cR}) \cosh (m_c {\cR}) - \sinh (m_c {\cR})}{(m_c {\cR}) \cosh (m_c {\cR}) + (m_b {\cR}) \sinh (m_c {\cR})}.
\end{equation}
\end{subequations}
Writing the solution for $\phi \simeq -\vi^2/12$ we finally get the solution
\begin{equation}
	\phi(r>{\cR}) \simeq \phi_b \left[1 - 2B\frac{\cR}{r}e^{-m_b (r-{\cR})}\right].\label{solution}
\end{equation}

\subsection{The thin shell effect}

The screening mechanism works well when the fifth force is suppressed by a physical mechanism, which means that the field turns null in desirables conditions (of density or  scale, for instance). The potential for a typical massive scalar field is a Yukawa potential, where the amplitude and the range depends on the environment properties. The thin shell effect takes place when the amplitude tends to zero in the neighbours of a compact object. It may happen in typical {\it chameleon} fields, whose solution around a sphere of constant density is
\begin{equation}
	\phi(r>\cR) = \phi_b  + (\phi_c-\phi_b)\frac{\cR}{r}e^{-m_b (r-\cR)}\;,
\end{equation}	
which is identical to the solution found in the previous section,  if one defines $\phi_c = \phi_b(1+2B)$. It means that we can also define a thin shell parameter in the $f(X)$ case as
\begin{equation}
	\frac{\Delta \cR}{\cR} = \frac{\phi_g - \phi_c}{6 \Phi_S} = -\frac{B}{3 \Phi_S}\phi_g\;.
\end{equation}
Therefore,
\begin{equation}
	\phi(r) = \phi_b + \frac{3}{4\pi}\frac{\Delta \cR}{\cR}\frac{M_c}{r}e^{-m_b (r-\cR)}\;.
\end{equation}
The required  condition, $\Delta \cR/\cR \ll 1$, is  satisfied for $B\ll 1$. According to the solution, it happens when $m_c \cR \ll 1$, and the amplitude may be approximated as
\begin{equation}
	B\simeq \frac{1}{3}m_c^2\cR^2 = \frac{1}{9}\rho_c \cR^2,
\end{equation}
which no longer depends on $m_b$. Thus,
\begin{equation}
	\frac{B}{3\Phi_S} = \frac{1}{f_R}\frac{\Delta \cR}{\cR} = \frac{1}{36\pi} \sim 10^{-2}\;,
\end{equation}
which  means that, for enough compact objects, the thin shell can  be approximated by the condition 
\begin{equation}
	\frac{\Delta \cR}{\cR} \simeq \frac{f_R}{36\pi}\;.
\end{equation}
In other words, this amounts to say that the screening  depends only upon the value of the field in the background. Using the values of the Sun ($\rho_\odot = 1.408 \mathrm{g\cdot cm^{-3}}$, $M=1M_\odot$ and $\cR=1\cR_\odot$),  we find a value much close to the predicted one, that is 
\begin{equation}
	\varepsilon_\odot = 1-\frac{36\pi}{f_{R\odot}}\frac{\Delta \cR_\odot}{\cR_\odot} \simeq 1.3\times 10^{-3}.
\end{equation}
Table \ref{planets_thinshell} shows the values of $\varepsilon$ for the solar system planets. In Fig. \ref{hists} we show how $\varepsilon$ is distributed for the solar system moons\footnote{https://www.wolframalpha.com/examples/SolarSystem.html} (A) and for stars which hosts exoplanets\footnote{http://exoplanets.org/} (B).  We can see that the screening conditions are satisfied in most of those objects, except for a very small fraction. It means that the previous assumptions, spherical symmetry and constant density, may be not valid or that the thin shell effect does not work. Therefore, these objects are very important in the study of modified gravity theories. Notably,  the solar system moons are the most promising cases, due to their proximity.

In next section we find some bounds for the background value of the field through the analysis of the thin shell effect in the Earth-Moon system.

\section{Astrophysical tests}

In order to test the viability of the $f(X)$ theory we analyze how it behaves in the neighbours of astrophysical  bodies, such as planets, moons, stars, etc. The field generated by these bodies can be described as small perturbations on the background value, which means that we can use the weak-field approximation. The perturbed metric, in the Jordan frame, is given by
\begin{equation}
	d s^2 = -[1-2\mathcal{A}(r)]dt^2 + [1+2\mathcal{B}(r)](dr^2+r^2d\Omega^2),
\end{equation}
where $\mathcal{A}$ and $\mathcal{B}$ are functions of $r$. The post-Newtonian parameter $\gamma = \mathcal{A}(r)/\mathcal{B}(r)$, in this context, is approximated to
\begin{equation}
	\gamma \simeq \frac{1-\Delta \cR/\cR}{1+\Delta \cR/\cR},
\end{equation}
provided that $m_b r \ll 1$, which is well satisfied in the $f(X)$ case. Here the mass of the field, $m_b = \sqrt{|f_{Rb}|/3f_{Rb}}$, is much smaller than in a metric $f(R)$, $m_b = 1/\sqrt{3f_{Rb}}$. Since $f_{Rb} \ll 1$.

\subsection{Solar system constraints}

The most direct bound that one can impose on $f(X)$ theories comes from the existence of the Earth atmosphere. The idea is that it can exist in a $f(X)$ gravity only if the thin-shell is smaller than the ratio between the atmosphere height and the Earth radius, i.e.
\begin{equation}
\frac{\Delta \cR_{atm}}{\cR_{atm}} < \frac{h_{atm}}{\cR_\oplus}\;.
\end{equation}
Using $h_{atm}\sim 10^{2}\mathrm{km}$ and $\cR_\oplus \simeq 6.3\times 10^{3}\mathrm{km}$ we find that $\Delta \cR_{atm}/\cR_{atm} < 1.6\times 10^{-2}$ and, therefore,
\begin{equation}
|f_{Rg}| < 1.8 \;.
\end{equation}
We also find a very similar bounds using exoplanets data. Following the method proposed in \cite{Santos2017} we find that $|f_{Rg}| < 2.6$.

Currently, the most restrictive measure of the deviations from the General Relativity is the one got by the Cassini Mission \cite{Bertotti:2003rm}. This mission provides data of light spectral deviation from gravity. The observed value indicates that gravity, inside the solar system, is in well agreement with the general relativity. The measured PPN parameter is $|\gamma_\odot-1| < 2.3\times 10^{-5}$ which gives the following constraints for the thin shell parameter
\begin{equation}
	\frac{\Delta \cR_\odot}{\cR_\odot} < 1.15\times 10^{-5}\;.
\end{equation}
Therefore,
\begin{equation}
	|f_{Rg}| < 1.3\times 10^{-3}\;.
\end{equation}

Finally, we find that the most stringent bounds comes from the imposition that the Earth-Moon system must remain bounded, these are the best constraints in a thin shell which we can reach so it should gives the best constraints on the background scalar field too. Such conditions can be expressed by the following inequality\begin{equation}
\eta = 2\frac{|a_{moon}-a_\oplus|}{a_{moon}+a_\oplus} < 10^{-13}\;, \label{eta}
\end{equation}
where $a_{moon}$ and $a_\oplus$ stand for the Moon and Earth accelerations, respectively (see \cite{CapTsu} for a discussion in the case of $f(R)$ gravity). In a $f(X)$ gravity scenario they depend directly on the Earth thin shell parameter, i.e.,
\begin{subequations}
\begin{equation}
	a_\oplus \simeq \frac{GM_\odot}{r^2}\left[ 1 + 3 \left(\frac{\Delta \cR_\oplus}{\cR_\oplus}\right)^2 \frac{\Phi_\oplus}{\Phi_\odot}\right]
\end{equation}
\begin{equation}
	a_{moon} \simeq \frac{GM_\odot}{r^2}\left[ 1 + 3 \left(\frac{\Delta \cR_\oplus}{\cR_\oplus}\right)^2 \frac{\Phi_\oplus^2}{\Phi_\odot\Phi_{moon}}\right]
\end{equation}
\end{subequations}
which gives the following value for the thin shell parameter
${\displaystyle \frac{\Delta \cR_\oplus}{\cR_\oplus} < 2\times 10^{-6}}$ 
or, equivalently,  
$|f_{Rg}| < 2.3\times 10^{-4}$.

\section{Conclusions}

The hybrid metric-Palatini $f(X)$ approach consists of the superposition of the metric Einstein-Hilbert action with an $f(\R)$ term constructed {\it \`a la} Palatini~\cite{Harko:2011nh,Capozziello:2015lza,Capozziello:2012ny}. In this work we have investigated the efficiency of the screening mechanism for this class of extended  gravity theories. We have computed the value of the field around spherical bodies embedded in a background of constant density and found that, under such conditions, the field is given by Eq. (\ref{solution}), whose solution depends only on the value of the  field at the background for most of the spherical self-gravitating objects, i.e., $\Delta \cR/\cR\approx f_{Rb}/36\pi$.

The viability of the model has been  evaluated comparing how the thin shell factor behaves in the neighborhood of different astrophysical objects, like planets and moons, such as the Sun and other stars which host planets. We find that the condition is very well satisfied except only for some peculiar objects, which may be important for future studies, mainly that ones close to us like the solar system moons.

We have also derived some bounds on the model using data from the solar system, such as the spectral deviation measured by the Cassini mission~\cite{Bertotti:2003rm,cassini}. The most stringent constraints comes from the condition (\ref{eta}), which is necessary to keep the Earth-Moon as a bounded system. It requires that the value of the field at the Galaxy background ($f_{Rg}$) must be less than $2.3 \times 10^{-4}$. We emphasize that the kind of analysis presented here helps understanding some additional properties of this class of theories out of the cosmological context, where they seem to provide  viable alternative to GR scenarios driven by the dark matter and dark energy fields.

\begin{acknowledgements}
MVS thanks the Brazilian founding agency CNPq for financial support. JSA acknowledges support from CNPq (Grants no. 310790/2014-0 and 400471/2014-0) and the Rio de Janeiro State agency FAPERJ (grant no. 204282). DFM thanks the Research Council of Norway for their support and the NOTUR cluster {\tt{HEXAGON}}, which is the computing facility at the University of Bergen. SC acknowledges the support of  the {\it Science Without Borders Program} - CNPq
No. 400471/2014-0
{\it Theoretical and Observational Aspects of Modified Gravity Theories}.
This paper is based upon work from COST action CA15117 (CANTATA), supported by COST (European Cooperation in Science and Technology).
\end{acknowledgements}

\end{document}